# Sound and Fury, Signifying Nothing? Impact of Data Breach Disclosure Laws

Muhammad Zia Hydari[*]    Yangfan Liang[†]    Rahul Telang[‡]

June 13, 2024

**Abstract**

Data breach disclosure (DBD) is presumed to improve firms' cybersecurity practices by inducing fear of subsequent revenue loss. This revenue loss, the theory goes, will occur if customers punish an offending firm by refusing to buy from them and is assumed to be the primary mechanism through which DBD laws will change firm behavior ex ante. However, our analysis of a large-scale data breach at a US retailer reveals no evidence of a decline in revenue. Using a difference-in-difference design on revenue data from 302 stores over a 20-week period around the breach disclosure, we found no evidence of a decline either across all stores or when sub-sampling by prior revenue size (to account for any heterogeneity in prior revenue size). Therefore, we posit that the presumed primary mechanism of DBD laws, and thus these laws may be ineffective and merely a lot of "sound and fury, signifying nothing."

[*]Katz Graduate School of Business, University of Pittsburgh, 3950 Roberto Clemente Dr, Pittsburgh, PA 15260, USA. `hydari@alum.mit.edu`. All correspondence about the paper should be sent to Zia Hydari.    [†]`yangfanl@alumni.cmu.edu`    [‡]Heinz College, Carnegie Mellon University, Hamburg Hall, Forbes Ave, Pittsburgh, PA 15213, USA. `rtelang@andrew.cmu.edu`

# Introduction

The large-scale adoption of information and communications technologies (ICT) and payment card systems in retail operations has led to increased efficiency and increased convenience for firms and consumers. For instance, consumers do not need to wait in lines and do not need to carry cash with them. Firms benefit too because of the efficiency of the transactions involving internet-connected payment card systems—not only is the exchange with the consumer more efficient (cf. cash transactions) but also the authorization process with the bank is faster and more reliable. Before the wide-spread adoption of internet-connected point of sale (POS) systems, retailers would use dial-up payment card systems that would require them to dial-up the payment processor's number for each transaction, transmit the payment card and purchase information, and then receive an approval or denial of the particular charge. The internet-connected POS systems not only increased operational efficiency but also enabled firms to collect real-time sales and marketing data. The advantages of these systems are undeniable but less discussed are the unintended disadvantages of these connected POS systems. One such disadvantage is that any device connected to the internet is a potential target for a cyber attack, which may emanate from anywhere in the world. These cyber attacks lead to data breaches, which impose a high cost on consumers.

A data breach is defined as the "breach of security leading to the accidental or unlawful destruction, loss, alteration, unauthorised disclosure of, or access to, personal data transmitted, stored or otherwise processed."[1] Despite decades of discussions to reduce the incidence of data breaches, these potentially adverse events continue to happen frequently. For instance, the California Department of Justice (CA DOJ) recently (circa 2022) made an authorized disclosure of the names, ages, and addresses of concealed-carry firearm permit applicants from 2011 to 2021, leading the state attorney general to "acknowledge the stress this may cause those individuals whose information was exposed."[13] These data breaches impose an externality on the data subjects whose information has been breached and happen not only at government entities such as the CA DOJ but at non-profit organizations such as hospitals with extremely sensitive health data, and at for-profit firms such as retailers.

In discussing a policy framework for data breaches, Telang [22] describes two ways in which policymakers regulate firms and organizations, viz., ex-post regulations which impose consequences on the firm after an adverse event has taken place, and ex-ante regulations which require firms and organizations to comply with requirements such as implementing specific technology standards before an adverse event has taken place. For instance, all US states (circa 2022) have passed laws that require firms to disclose data breaches to the data subjects whose information may have been breached. Telang [22] classifies these disclosure laws as ex-post regulation as they impose transparency requirements on the firms about the data breach after the adverse event has occurred. Telang [22] further states that "little empirical evidence supports or disputes disclosure laws' effectiveness."[1] This lack of empirical evidence persists and the question on whether these data breach disclosure laws are effective is far from settled.

Data breach disclosure laws (DBDL) are purported to force firms to take corrective actions to reduce any externalities through two mechanisms: first, the primary mechanism through which DBDL will propel these firms to act is to protect their reputation and the demand for their goods and services, and second, firms will act to protect themselves from common law tort liability. A plausible additional advantage of DBDL is that it may allow consumers to take protective measures to mitigate the risk arising from the data breach.

---

[1] See also Romanosky, Telang, and Acquisti [18].

In this paper, we empirically investigate the primary DBDL mechanism which has been theorized to lead to firms' improving their security and reducing the incidence of data breaches, i.e., the fear of loosing firm reputation and a reduction in the demand for firm's goods and services. Specifically, we examine the impact of a large data breach at the largest US home improvement retailer, Home Depot, on the subsequent demand observed at its stores. If the proposed primary mechanism is present, it would lead to a (ceteris paribus) reduction in demand for the firm whose data breach was disclosed. Alternatively, if no empirical evidence is observed for the primary mechanism, it would cast doubts on the effectiveness of data breach disclosure laws. Contrary to the theoretical claims, we find no evidence of a decline in revenue at Home Depot after data breach disclosure. Therefore, we posit that the presumed primary mechanism of DBD laws, and thus these laws may be ineffective and merely a lot of "sound and fury, signifying nothing."

## Literature Review

As the information security literature is vast, we focus our attention on recent research that relates to data breaches. One theme in this literature is the information security risk that arises from vulnerabilities and threats to organizations. Johnson [9] analyzed the risk of inadvertent disclosure through P2P file-sharing networks, revealing significant vulnerabilities in large financial institutions. He found a correlation between information leakage and firm characteristics such as the employment base and number of retail accounts. Extending this discussion, Kwon and Johnson [10] explored how security performance and compliance interact in healthcare settings, emphasizing the role of organizational maturity in managing these risks. Relatedly, Liu et al. [12] investigated the impact of centralized IT decision-making on cybersecurity breaches in higher education institutions, finding that centralized IT governance is associated with fewer breaches. Furthermore, Zhuang et al. [24] examined the impact of informing firms about their security vulnerability. They found that such awareness led firms to improve their security.

In addition to identifying risks, understanding organizational and customer responses to security breaches is crucial. Choi et al. [5] tested a model to understand how a firm's incident response reduces the impact of a breach on their relationship with customers. Janakiraman et al. [8] studied the impact of a data breach announcement for a multichannel retailer, in which customers in only one channel were breached. They found that spending by customers in the affected channel declined, however the effects were less pronounced on customers with high prior patronage. Moreover, breached customers switched to the unaffected channel, leading the authors to suggest that firms should invest in multiple channels. Finally, the authors find that any negative effect on customer spending wanes over time. Gwebu et al. [6] studied the role of corporate reputation and crisis response strategies in mitigating the financial impacts of data breaches. They found that high-reputation firms are less dependent on response strategies, but only a few response strategies worked for lower-reputation firms. Additionally, Ng et al. [14] offers guidance on creating impactful fear appeals that prevent attitudinal ambivalence and thereby promote the adoption of security protection behaviors.

Another theme is the relationship between IT strategies, security investments, and security breaches. For instance, Kwon and Johnson [10] examined how operational security maturity influences how breaches impact compliance in healthcare. Li et al. [11] found that IT security investments reduce breaches in less digitalized organizations but surprisingly increase them for highly digitalized organizations. These highly digitalized organizations attract sophisticated hackers who extract hints from IT security investments when selecting targets. Finally, Wang et al. [23] found that greater IT innovativeness is associated with a higher risk of data breaches, but several factors, such as managerial IT expertise, board connections with cybersecurity managers, and environmental complexity, can mitigate or amplify this relationship.

While the extant literature has examined several issues around breaches, our goal is to examine the efficacy of data breach disclosure laws through their purported primary mechanism of revenue loss for the firm. Thus, our paper attempts to estimate the broader impact a data breach disclosure has on the firm's revenue at the aggregate store level. We can then relate it to the effectiveness or lack thereof of data breach disclosure laws.

## Contextual Background

For our empirical analysis, we study a large-scale data breach in the year 2014 at Home Depot, the largest US home improvement retailer. Home Depot operated 1,977 physical stores in the US and its territories in the fiscal year ending in February 2015, hereafter fiscal 2014. These US stores were located in all 50 US states, the District of Columbia, Puerto Rico, the Virgin Islands, and Guam.[2] Home Depot's revenue in fiscal 2014 was approximately 83 billion US dollars, with in-store sales accounting for 95.5 percent of the total revenue or roughly 80 billion US dollars.[3] Home Depot accepted multiple modes of payment, including credit card and debit cards and considered these electronic payment modes as crucial to its business operations.[4] In September 2014, Home Depot acknowledged a massive data breach of its customers' payment card data. According to Home Depot's own estimate, cyber criminals stole 56 million credit card and debit card numbers over a several month period, making it the largest retail breach ever, circa 2014. Although the breach itself was bad enough, the situation was compounded by the fact that Home Depot appeared to be unaware of the breach for several months. It was not until September 2nd, 2014, when the breach was first disclosed by an independent cyber journalist, Brian Krebs, that the consumers and apparently Home Depot became aware of the breach.

---

[2] Please see https://thequarterly.org/sec-filings/hd/2015/10-k.html   [3] Home Depot reported 4.5% sales were through the online channel in fiscal 2014. Although growing, Home Depot's online sales were insignificant compared to its in-store sales, as its product assortment is less suitable for online sales. Please see https://thequarterly.org/sec-filings/hd/2015/10k.html   [4] ibid.

The Home Depot 2014 data breach was reportedly caused by a variant of a malware known as "BlackPOS," which was installed on Home Depot's store registers.[5] The malware was able to steal payment card information, including cardholder names, card numbers, expiration dates, and security codes, from customers who used their cards during in-store transactions. Home Depot later disclosed that hackers penetrated into its computer network using a vendor's stolen login information, and having once penetrated into the network, were subsequently able to install the BlackPOS variant malware on store registers. The malware was removed by November 2014 according to Home Depot.[6] In a press release dated September 18th, 2014, Home Depot warned the public and investors about the impact of the breach on its financial results. The listed financial risks alluded to the costs of liabilities, litigation, and remediation. Cru-cially, Home Depot did not speculate any risk to future revenue or of losing its affected customers as a result of the breach of their payment card data. Despite theoretical claims that data breaches should lead to revenue loss due to customer punishment, Home Depot either did not anticipate any impact or did not think it was a significant enough concern to mention. We empirically examine if Home Depot stores experienced any revenue loss due to this massive data breach.

---

[5] https://krebsonsecurity.com/2014/09/home-depot-hit-by-same-malware-as-target/

[6] https://www.usatoday.com/story/money/business/2014/11/06/home-depot-hackers-stolen-data/18613167/

## Data, Identification, and Model

### Data

To study the impact of this data breach, we use a unique dataset which provides us with all debit card transactions at Home Depot stores processed through one of the largest US bank. While our non-disclosure agreement prohibits us from disclosing the identity of this bank and the states that it operates in, we were able to obtain Home Depot sales data for 302 stores spanning five US geographical states in the Middle Atlantic, South Atlantic, and Midwest regions. Assuming that the customers primarily reside in the same state as the store's location, these Home Depot stores served a combined population of approximately 50 million people in 2014. The data span a 20-week period around breach disclosure at Home Depot. Specifically, we obtained data from June 30, 2014 to November 16, 2014, which gives us week 1 to 9 pre-breach period, week 10 for the data breach disclosure (DBD), and week 10 to 20 for the post-DBD period. The sales data provided to us is at the individual transaction level and notably includes store information, transaction amount, and transaction date. We aggregated these data to calculate the weekly sale, in thousands of US dollars, at the store level. The natural log of store sales is our primary outcome variable. We also aggregated the weekly count of customer visits to these Home Depot stores as a secondary outcome.

In the US, Home Depot's primary competitor is Lowe's Companies Incorporated (Lowe's). Given the product assortment these stores carry, the Home Depot and Lowe's competition is between their brick-and-mortar stores with the online channels being less important, especially during the time period of the study. As we will elaborate later, the competition faced by Home Depot stores is a crucial aspect of our analysis. To measure the competition between Home Depot stores in our sample, we calculated the distance between a Home Depot and its closest Lowe's store. We carried out this distance calculation by collating the latitude and longitude information of the Home Depot and Lowe's store from publicly available geographical data. This distance is compared to a threshold to determine if a focal Home Depot store faces any competition.

## Identification and Model

Our goal is to estimate the impact of the data breach disclosure at Home Depot (HD) on subsequent short-run sales to examine the theoretical supposition that data breach disclosures (DBD) would lead to a decline in sales. To identify this impact, an obviously infeasible ideal would be to run a randomized controlled trial (RCT) in which data breach and its subsequent disclosure were randomized onto HD stores. Given the infeasibility of an RCT, we resort to an observational study but even the observational data does not lead to a straightforward empirical setup. To elaborate, all HD stores are simultaneously affected by the breach so we do not have a straightforward control group from amongst the HD stores. A before-after design, also known as interrupted time-series, is possible but the defects of such a design in our setup, which is affected by seasonality, would render any estimates implausible. Moreover, Lowe's stores, while not directly breached, cannot act as the control stores as they serve as competition to HD and some if not all potential sales lost by HD will likely be gained by Lowe's as angry customers may substitute HD with Lowe's. This potential spillover sales would be a stable unit treatment value assumption (SUTVA) violation [7]. While the competitive position of Lowe's precludes the direct use of Lowe's stores as controls, nonetheless, we are able to exploit the presence of Lowe's stores to construct an ersatz control group. Consider a HD store with no Lowe's store in its vicinity, which makes it a monopoly in its geographical market. Given the product assortments that home improvement stores carry, many if not all customers will grudgingly continue to patronize this HD store because of lack of alternatives. Thus, the HD store sales in monopoly markets are likely to have little to no impact in the short run subsequent to a DBD. In contrast, customers of HD stores with Lowe's stores in their vicinity, have a realistic alternative so they may actually substitute their supplier from HD to Lowe's in response to a HD DBD. While all HD in the US suffered from the data breach, the ability of a focal store's customer base to punish HD is constrained by the availability of a Lowe's store within its vicinity. We exploit this feature of our study context to construct both a treatment group and a control group from the HD stores in our sample. We posit that a focal HD store with a Lowe's store within a 3-mile radius has a credible competitor and hence may face a reduction in revenue after this DBD. In contrast, a focal HD store with no Lowe's store in the vicinity is less likely to lose customers and its revenue. Thus, we use the former set of stores as the treatment group and the later set of stores as the control group. This setup depends on the assumption that a 3-mile radius threshold reasonably classifies HD stores into monopoly (control) or competitive (treatment) stores. The rationale for this distance measure is that a Lowe's store that is 3-mile away from a competing HD store would, in most cases, add less than 15 minutes driving time to a potential buying trip, which imposes a small enough additional cost on a buyer that it may not thwart them to switch from HD to Lowes. While our main results are based on the 3-mile radius, we analyze with other radius thresholds as a robustness check of our inference. Finally, we also use a continuous distance measure for competition as an additional robustness check.

The model for our analysis is a standard difference-in-difference (DID) specification:

$$Y_{it} = \beta_0 + \beta_1 \mathbb{1}(d_i \leq x) \times POST_t + \theta_i + \gamma_t + \epsilon_{it} \tag{1}$$

$Y_{it}$ is the outcome variable (primarily the natural log of store sales but also log of the number of visits, store sales, or the number of visits) measured at store $i$ during week $t$. The presence of competition is indicated by the distance between the focal HD store and its nearest Lowe's store, $d_i$ being less than or equal to a particular radius threshold $x$. So, HD stores which do not have a close by Lowe's store relative to the threshold $x$ are deemed to be a geographical monopoly and evaluate a value of 0 for this indicator variable, $\mathbb{1}(d_i \leq x)$. $POST_t$ is an indicator if the observation is from a post DBD period (i.e., $t \geq 10$). $\theta_i$ and $\gamma_t$ are the usual unit and time fixed-effects used in a DID design. A particular HD store, $i$, is deemed to be treated in week $t$ if $\mathbb{1}(d_i \leq x) \times POST_t = 1$. The DID design depends on the parallel trends assumptions, i.e., the monopoly store and competitive stores will have similar time trends in the absence of the treatment. We will probe this assumption using event-study charts (also known as lags and leads plots) in a later section. A recently publicized concern with DID models is that estimation may be biased in case of staggered adoption of treatment [3]. In our study context however, all stores are subject to DBD at the same time so potential bias from staggered adoption is not a concern.

# Results

Table 1: Main Effect of Data Breach

| Dependent Variables: | sales | logSales | visits | logVisits |
| --- | --- | --- | --- | --- |
| Model: | (1) | (2) | (3) | (4) |
| *Variables* | | | | |
| POST × TREATd | -20.77 | -0.00 | -0.31 | 0.01 |
| | (48.99) | (0.09) | (0.35) | (0.03) |
| Threshold Distance | 3 | 3 | 3 | 3 |
| Control Stores | 114 | 114 | 114 | 114 |
| Treated Stores | 188 | 188 | 188 | 188 |
| *Fixed-effects* | | | | |
| hdid (302) | Yes | Yes | Yes | Yes |
| week (20) | Yes | Yes | Yes | Yes |
| *Fit statistics* | | | | |
| Observations | 6,040 | 6,040 | 6,040 | 6,040 |
| $R^2$ | 0.75 | 0.53 | 0.94 | 0.78 |

One-way (hdid) standard-errors in parentheses
Signif. Codes: ***: 0.01, **: 0.05, *: 0.1

Table 1 shows the estimation results for specification (1) for the four outcomes,

store sales (in thousands of US dollars) and its natural log, and number of visits and its natural log, using 3-miles as the radius threshold for competition (i.e., $x = 3$ for specification 1). For all outcomes, the results are small in magnitude and statistically not significant. For instance, the point estimate for the treatment effect on natural log of store sales is approximately 0. Since the treatment coefficient in the log-linear model has a semi-elasticity interpretation, this suggest no-effect on the outcome. Moreover, this point estimate is not statistically significant.

To probe the parallel trends assumption, we first explored raw outcome plots. Next, we estimated an event-study model using the following specification:

$$Y_{it} = \beta_0 + \sum_{\substack{\tau=1 \\ \tau \neq 9}}^{20} \beta_\tau \mathbb{1}(d_i \leq x) \times \mathbb{1}(t = \tau) + \theta_i + \gamma_t + \epsilon_{it} \qquad (2)$$

Here, we use relative time interacted with the competition classifier $\mathbb{1}(d_i \leq x)$ to simulate placebo treatment indicators in the pre-treatment period as well as specifying a separate treatment indicators for each post-treatment period. To avoid the dummy variable trap, we drop the indicator for week 9, i.e., the period immediately before the treatment period. We estimated this model and plotted the $\beta_\tau$ coefficients, which represent the placebo and actual effect of the treatment on the particular outcome (i.e., store sales or store visits).

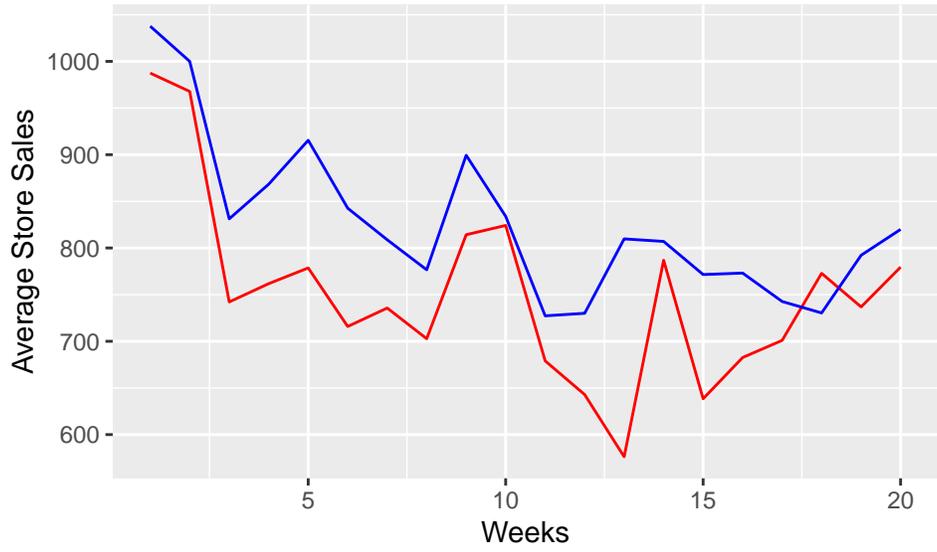

Figure 1: Store Sales over Time (competitor withing 3 miles)

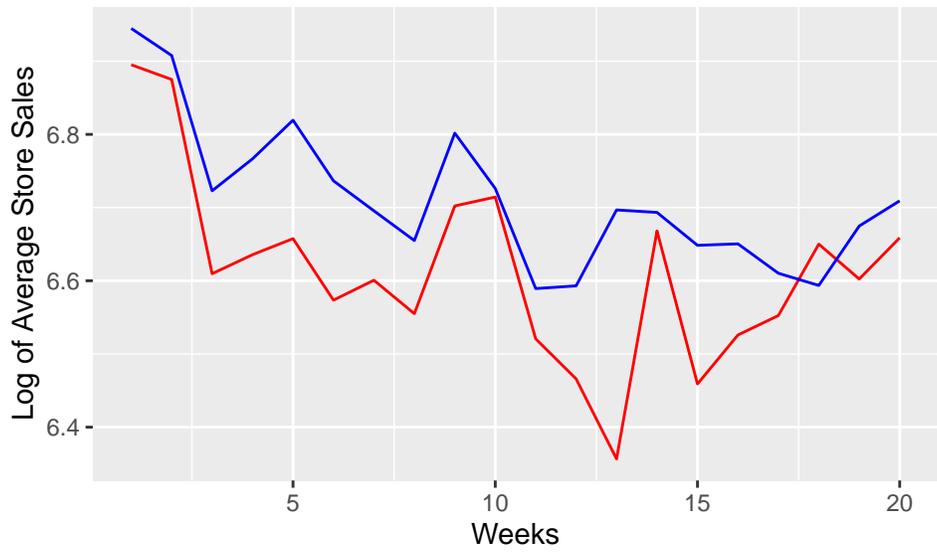

Figure 2: Log of Store Sales over Time (competitor withing 3 miles)

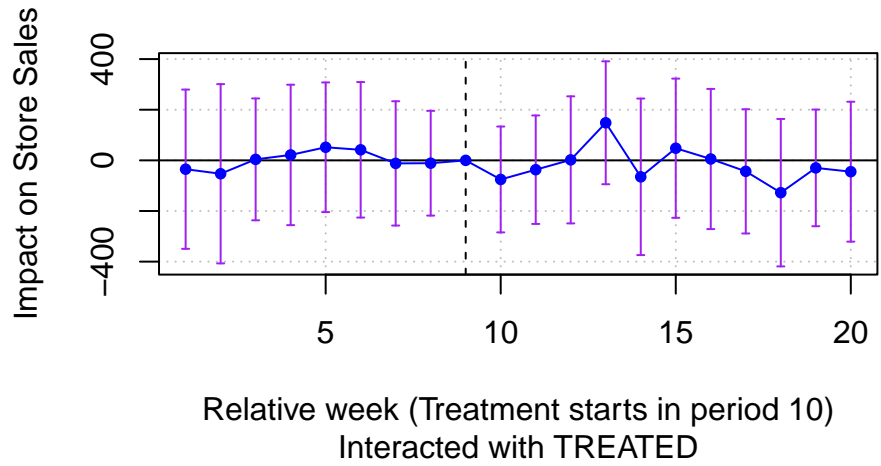

Figure 3: Dynamic Effect of Data Breach on Store Sales (competitor within 3 miles)

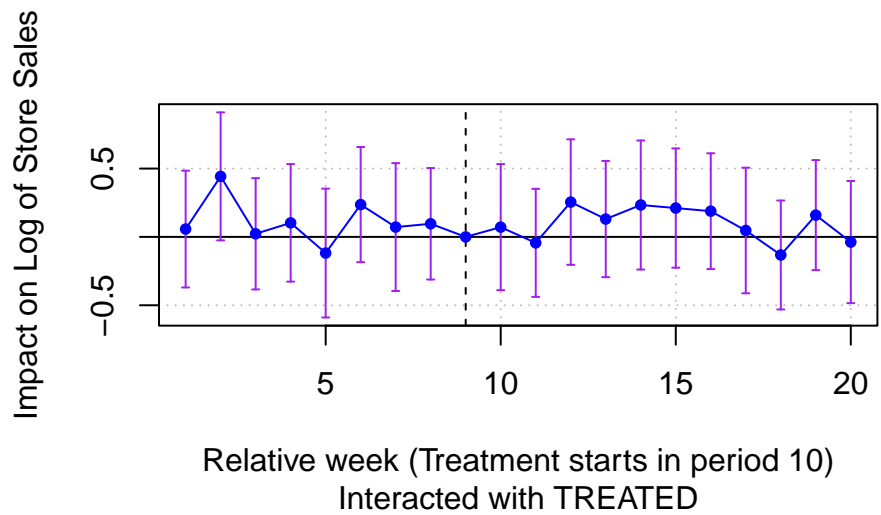

Figure 4: Dynamic Effect of Data Breach on Log of Store Sales (competitor within 3 miles)

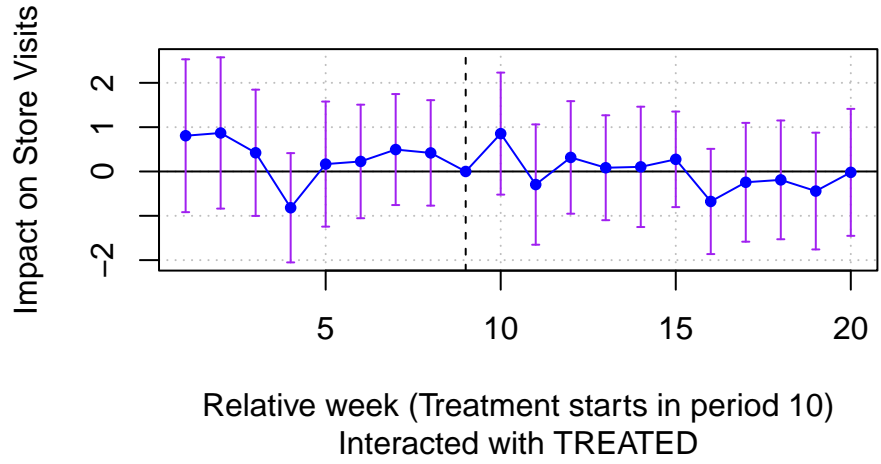

Figure 5: Dynamic Effect of Data Breach on Store Visits (competitor within 3 miles)

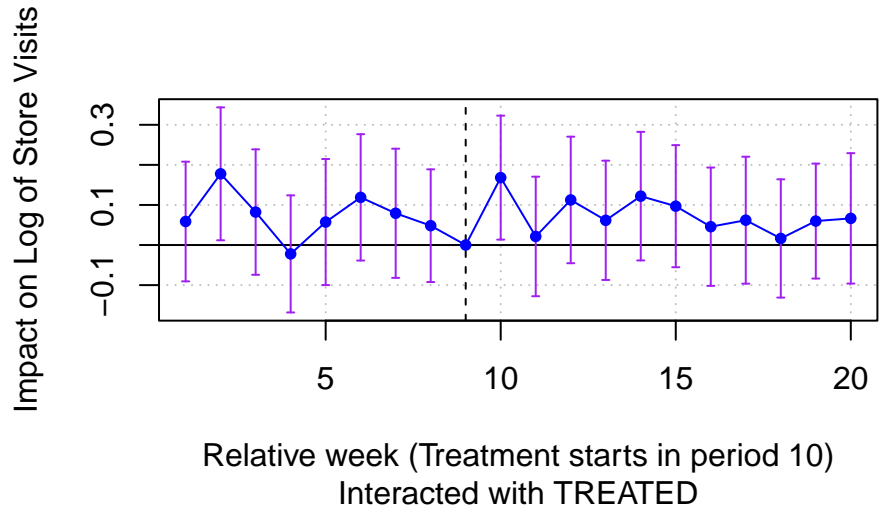

Figure 6: Dynamic Effect of Data Breach on Log of Store Visits (competitor within 3 miles)

Figures 1 and 2 plot the average sales and natural log of average sales for the treated and control stores. These sales measures for the two groups, while not strictly parallel prior to any regression adjustments, nonetheless move in the same direction in the pre-treatment period (except for one period). To check for the presence of pre-treatment differential trends between the treated and control stores after adjustment, we plotted event study diagrams (also known as leads and lags plots) as is common in the difference-in-difference literature. Figures 3 and 4 show these event study charts for store sales and log of store sales respectively. In both cases, the point estimates in the pre-treatment periods are null, arguing for the absence of any differentials trends across the treatment and control groups. This suggests that the stores in our treatment and control groups are comparable. Additionally, the post-treatment estimates are also null, indicating the absence of any effect from DBD on store sales or log of store sales. We also estimated and plotted event study charts for average visits and log of average visits and these charts are shown in Figures 5 and 6. For these outcomes too, there are no discernible pre-treatment trends. Additionally, the post-treatment estimate are all null.

# Robustness

## Varying Radius Threshold to Establish Competition

For our main models, we used a 3-miles radius threshold to classify HD stores into treatment and control groups. We motivated this choice from a qualitative argument that this 3-mile distance would approximately be less than 15 minutes of additional driving time for the buyer, and thus a small enough cost for the buyer to not thwart them from switching to the competitor. While plausible, this threshold neither has a strong theoretical underpinning nor is it empirically established. Thus, we carried out a sensitivity analysis of our main results by choosing different radius thresholds and re-estimating specification 1.

Specifically, Table 2 presents the estimation results for radius thresholds of $x$ miles, where $x \in \{1, 2, 4, 5, 6\}$, in columns 1–5 respectively. The estimates for the treatment effects are null, which strengthens the credibility of our main estimates.

## Varying Treatment Intensity Using Continuous Distance Measures

In our main analysis, we dichotomize our treatment by choosing a radius threshold value of 3-miles for classifying the HD stores into monopoly and competitive stores. While

Table 2: Main Effect of Data Breach Various Distance

| Dependent Variable: | logSales | | | | |
|---|---|---|---|---|---|
| Model: | (1) | (2) | (3) | (4) | (5) |
| Variables | | | | | |
| POST × TREATd | 0.03 | -0.04 | 0.02 | 0.04 | 0.01 |
| | (0.08) | (0.08) | (0.09) | (0.11) | (0.13) |
| Threshold Distance | 1 | 2 | 4 | 5 | 6 |
| Control Stores | 188 | 148 | 88 | 63 | 52 |
| Treated Stores | 114 | 154 | 214 | 239 | 250 |
| Fixed-effects | | | | | |
| hdid (302) | Yes | Yes | Yes | Yes | Yes |
| week (20) | Yes | Yes | Yes | Yes | Yes |
| Fit statistics | | | | | |
| Observations | 6,040 | 6,040 | 6,040 | 6,040 | 6,040 |
| $R^2$ | 0.53 | 0.53 | 0.53 | 0.53 | 0.53 |

One-way (hdid) standard-errors in parentheses
Signif. Codes: ***: 0.01, **: 0.05, *: 0.1

we further tested our analysis by using different radius threshold values, an alternative approach would have been to use the actual distance between the focal HD store and the competing Lowe's store rather than the dichotomous competition indicator. The specification for such a model would be:

$$Y_{it} = \beta_0 + \beta_1(d_i \times POST_t) + \theta_i + \gamma_t + \epsilon_{it} \qquad (3)$$

Specification 3 is similar to a dose-response model and could potentially provide a nice marginal effects (aka causal response) interpretation while avoiding the somewhat subjective radius threshold choices. However, such continuous treatment models are notoriously hard to estimate in observational data as identification requires very strong assumptions,[4] which in turn reduces the credibility of the estimates. However, our goal here is to present this model as a robustness check as-if the identifying condition are met.

Table 3 presents the result of this continuous treatment model. The point estimate is 0 (up to two significant digits). Moreover, this point estimate is statistically not significant.

Table 3: Effect of Data Breach (Using a Continuous Measure)

| Dependent Variable: | logSales |
|---|---|
| Model: | (1) |
| *Variables* | |
| minDistMiles × POST | 0.00 |
|  | (0.01) |
| *Fixed-effects* | |
| hdid (302) | Yes |
| week (20) | Yes |
| *Fit statistics* | |
| Observations | 6,040 |
| $R^2$ | 0.53 |

One-way (hdid) standard-errors in parentheses
Signif. Codes: ***: 0.01, **: 0.05, *: 0.1

# Heterogeneity Across Store Based on Prior Sales

Although we have hitherto found null effects for DBD, it is plausible that the full sample average may be masking heterogeneity across store types. For instance, the DBD may affect stores with low sales differently than the stores with high sales.

To examine this hypothesis, we first divided our sample by the four quartiles. We estimated specification 1 with threshold distance set to 3-miles for these sub-samples, i.e., a model similar to our main model. Table 4 shows the result of this analysis. Again, there is no evidence to suggest that there is a negative impact on the sales for the stores, sub-sampled by their prior sales quartiles.

Table 4: Effect of Data Breach (By Prior Store Sales)

| Dependent Variable: | logSales | | | |
|---|---|---|---|---|
| Model: | (1) | (2) | (3) | (4) |
| Variables | | | | |
| POST × TREATED | -0.01 | -0.09 | 0.13 | 0.00 |
| | (0.15) | (0.10) | (0.19) | (0.23) |
| Quartiles | 4 | 3 | 2 | 1 |
| Fixed-effects | | | | |
| hdid | Yes | Yes | Yes | Yes |
| week (20) | Yes | Yes | Yes | Yes |
| Fit statistics | | | | |
| # hdid | 76 | 75 | 75 | 76 |
| Observations | 1,520 | 1,500 | 1,500 | 1,520 |
| $R^2$ | 0.64 | 0.36 | 0.34 | 0.35 |

One-way (hdid) standard-errors in parentheses
Signif. Codes: ***: 0.01, **: 0.05, *: 0.1

## Summary of Empirical Results

From our analysis of store sales both in aggregate and divided into sub-samples by prior revenue, we do not find any statistically significant evidence of a decline in store revenues. This lack of evidence suggests that the purported mechanism—i.e., a decline in revenue because of a data breach disclosure will lead firms to correct their cyber-security behavior—is not supported by empirical evidence. Hence, the expected improvement in cyber-security behavior of the firm from data breach disclosure is not likely to materialize if there is no impact on revenue.

# Discussion

The presumed primary mechanism for data breach disclosure laws (DBDL) is a reduced demand for goods and services of a firm that had a data breach. This mechanism is also posited to be the primary mechanism for free market unregulated approaches to the problem of data breaches. The intent of the article is to draw attention of the information systems community to the need of generating empirical evidence on whether the purported primary mechanism for data breach disclosure legislation is working. In our empirical analysis, we are unable to find any evidence of reduced demand for product and services in our investigation of a large scale breach at a major US retailer. Thus, if the DBDL are presumed to improve cybersecurity through an ex-post reduction in the offending firm's revenue, the absence of revenue reduction is unlikely to lead firms to improve their cybersecurity practices. Indeed, the BlackPOS malware whose variant led to the Home Depot data breach was also used in an earlier high-profile data breach at Target Corporation in 2013. The subsequent disclosures apparently did not spur Home Depot into improving their cybersecurity practices.

While not without limitations, our study is the only one we know that frames and then empirically addresses the primary mechanism of DBDL, in an area where perfect data is hard to come by. More importantly, our study calls for more research in interventions that may help ameliorate the status quo in which there are frequent data breaches, which impose small externality costs (such as increased monitoring on a very large number of people) and large externality costs on a small number of people. As Solow-Niederman [20] states, "individuals are left, at best, in a state of data insecurity and, at worst, in a compromised economic situation." Although the societal costs, taken together, are high, the collective response does not impose enough demand reductions on the liable firms, and in many cases negligently liable firms, for them to change behavior on their own.

Although our empirical investigation is focused on the purported primary mechanism of DBDL, i.e., a reduction in demand for product and services leading firms to take corrective action, we will briefly comment on the purported second mechanism, i.e., corrections induced by the possibility of common law tort liability. As Peters [17] opines, "*most importantly, even when these various state data-breach laws are effective and consumers are notified of a breach, they have almost no legal recourse against the entity whose security breach led to the unlawful or unauthorized procurement of their personal information. There is no clear-cut state or federal civil cause of action for consumers to bring, and existing causes of action have had limited success when applied to data breaches due to issues with standing and injury. Therefore, a stronger data-breach notification regime that provides consumers with a remedy when a data breach does occur and that is more effective in preventing data breaches from happening should be considered.*" Alicia Solow-Niederman, a legal scholars, has suggested a common law solution to the problem, while qualitatively discounting DBDL and ex ante technology standards. Specifically, Solow-Niederman [20] proposes that "courts can reinvigorate the tort of breach of confidence as a remedy for aggrieved consumers." Our intention here is not to evaluate this specific proposal but merely to point out that myriad research communities consider this problem as unsolved and looking for a solution.

Federal legal remedy in privacy lawsuits is suspect as the Supreme Court has recently dismissed lawsuits for lack of standing as discussed by Electronic Privacy Information

Center.[7] As Park [15] states: *Although a firm's misaligned incentive to invest in security measures is basically an agency problem to be addressed by data breach litigation, the U.S. courts' reluctance to grant Article III standing has reduced potential plain-tiffs' chance of winning and propensity to litigate, impairing the functionality of the private enforcement.*

Recent state-level legislative action in the US that goes beyond DBDL such as the California Privacy Rights Act (CPRA) may be promising. However, CPRA became effective on January 1, 2023 so its impact would need to be evaluated subsequently. Other US states have DBDL, which we have analyzed empirically in this paper. Thus, the problem of data breaches is still open to solutions. We reiterate two well-known difficulties in formulating a policy solution to this problem: (i) it is infeasible to eliminate data breaches entirely, which necessarily entails a balancing act of trading off data breaches versus the ease of operations, et cetera, (ii) the policy solution should be able to handle the dynamism of this area such that when tactics are evolved by the adversaries of the policy (e.g., legitimate adversaries finding loopholes, or illicit adversaries) the policy is able to adapt accordingly.

As we alluded earlier in the context of discussing Telang [22], there are two policy mechanisms to deal with data breaches, i.e., ex ante regulations that require adoption of technology standards and safeguards to prevent some or most data breaches or ex post regulations that hold firms liable for data breaches once they have occurred. The third option is to let the "invisible hand" of the free market induce firms to correct their behavior.

---

[7] https://epic.org/issues/consumer-privacy/article-iii-standing/

A defense of DBDL could be based on the argument that it solve the information problem for the consumers. For instance, Shapiro [19] states: "*Consumer protection regulation is based on a belief that the private market fails in a significant way: it fails to supply consumers with adequate information for them to make efficient choices among products and to protect themselves from unscrupulous sellers. Consumer protection regulation would be unnecessary in a world of perfect information. It would also be unnecessary if markets for information worked perfectly.*" We contend that DBDL do not solve the perfect information problem. The way to think about the issue is the following: consider a consumer who is about to engage in a transaction with a retailer such as Home Depot. In the context of a data breach, "perfect information" would mean that the consumer knows a priori that the retailer has not made sufficient investments in cybersecurity so that confidential data (e.g., debit card number) for that particular transaction would be disclosed to a cyber-criminal. In addition, the consumer knows all the ways that the cyber-criminal could abuse the confidential data. The consumer also knows and understand how the cyber-criminals may collate information from various breaches and the ensuing harm that occur. While we do not have empirical evidence to adduce, but anecdotally we have not found any consumer who would be willing to engage in a transaction for which the consumer knows that it would be breached. Yet, this "perfect information" is not available to consumers when they are engaging in transactions with businesses. To compound the problem, even if the consumer knew that a particular firm suffered from a breach, the marginal increase in the consumers' subjective probability of future breaches is so small as not to dampen demand.

The firm's data security state is latent from the consumer. In fact, given the high complexity of computer systems, even a particular firm that has "least cost access to the information," [19] may actually not know their own data security state because of underinvestment in that function.

There are several plausible criticisms that proponents of unregulated free-market market regimes can make to point to either the futility or the lack of need of regulations. One plausible example for the latter in our context that can be adduced by the free-market proponents is the payment card industry data security standard (PCI DSS). The PCI DSS is an industry standard that seemingly came into existence without a direct coercive role of the government and imposes technology standard on firms that desire to receive payments from their customers through the use of payments cards. The free-market proponent can use this example to argue that market mechanisms, without any government coercion or intervention, led to the creation of PCI DSS. Telang [22], who is not arguing for free-market unregulated regimes, correctly concedes that "the PCI standard is a proposed self-regulation rather than a government mandate," but does not necessarily say that PCI DSS came into existence without any role of the government. Contrary to the superficial appearance, it can be argued strongly that PCI DSS came into existence because of federal laws such as the Electronic Funds Transfer Act, common law such as Judd vs. Citibank, and regulations, that put financial responsibility on the payment card industry and banks in case of disputes. For instance, 15 U.S. Code §1643 states "**Burden of proof**: *In any action by a card issuer to enforce liability for the use of a credit card, the burden of proof is upon the card issuer to show that the use was authorized or, if the use was unauthorized, then the burden of proof is upon the card issuer to show that the conditions of liability for the unauthorized use of a credit card, as set forth in subsection (a), have been met.*"[8] These US laws can plausibly be claimed to be the filip that led to the industry improving their security and even creating a standard such as PCI DSS to improve downstream firms' security. The behavior of British bankers a few decades ago in the absence of such consumer protection is instructive. Anderson [2] asserts that British banks (circa 1994) would assert that their systems are infallible and argue that the consumer was responsible for any disputed transactions. Anderson [2] also argues that these British banks, unregulated or less regulated compared with US banks with respect to fraudulent transactions and system security, were spending more on security equipment but their objective was not security but "due diligence." The US banks, who were liable for fraudulent transactions, actually cared about security and achieved better security outcomes with less spending then their British counterparts.

---

[8] https://www.law.cornell.edu/uscode/text/15/1643

One classic criticism on regulations is that of regulatory capture [21]. While this concern is certainly valid and needs to be guarded against, it does not necessarily follow that an unregulated approach is better. If data breach or data privacy regulations were indeed beneficial to the producers in the industry, it does not explain the opposition to such regulations by industry groups. For instance, the opposition to California Consumer Privacy Act (CCPA) by various industry groups.[9] The proponents of non-regulation may speciously object that these companies have been able to get their way through lobbying or through finding loopholes but the fundamental question remains: if "regulatory capture" enables industry to make use of regulatory bodies to work in their favor, why would industrial lobby groups of self-interested firms oppose them?

Another well-known criticism of regulation is offsetting responses,[16] in which agents that seemingly benefit from a regulation resort to offsetting behavior that ends up not improving the desired outcome. To be specific, Peltzman [16] claimed that seat belt laws have not reduced overall mortality because of offsetting effects. While the empirical analyses in that work are questionable,[10] even granting that offsetting effects may exist in specific cases, it is hard to think of such scenarios in the context of firm's breaching consumer data and imposing an externality on the consumers. Please consider: in the case of automobile safety, the safety regulation applies to the automobile firm whereas the offsetting response is presumed to be from the consumer. That is, product that is safer due to safety regulation is then controlled and operated by the consumer. The claimed offsetting behavior is that the consumer resorts to more riskier driving because of an increased sense of safety and protection, with a plausible unintended consequence on the desired outcome, i.e., mortality. Thus, the claim of offsetting behavior is at least theoretically plausible in the case of automobile safety. In the case of data breaches caused by firms that impose an externality on the consumers, the data remains under the operational control of the firms whether it is safely maintained or not. Plausible scenarios in which consumer resort to risky behavior as a result of consumer protecting data breach regulation that increase the likelihood or magnitude of the data breach are hard to imagine.

---

[9] https://www.cpomagazine.com/data-protection/google-other-tech-companies-trying-to-dilute-ccpa-with

In conclusion, our findings challenge the theoretical claims that data breach disclosure laws (DBDL) lead firms to enhance their cybersecurity practices through the fear of revenue loss. Our empirical analysis of Home Depot's 2014 data breach reveals no significant impact on the company's revenue, suggesting that the anticipated customer punishment mechanism may be ineffective. This raises important questions about the efficacy of DBDL in their current form. Policymakers should consider alternative or supplementary strategies, such as ex-ante regulations, incentives for proactive cybersecurity investments, or stronger legal frameworks to hold firms accountable.

---

[10] See Sam Peltzman's interview on https://www.chicagobooth.edu/review/sam-peltzman-thinks-you-should-belt